\documentclass[conference]{IEEEtran}
\IEEEoverridecommandlockouts
\usepackage[T1]{fontenc}
\usepackage{amsmath}
\usepackage{amsfonts}
\usepackage{graphicx}
\usepackage{color}

\usepackage{amsmath}
\usepackage{cite}
\usepackage{amsfonts}
\usepackage{amssymb}
\usepackage{graphicx}
\usepackage{subfigure}
\usepackage{algorithm}
\usepackage{algpseudocode}
\usepackage{mathtools}

\usepackage{textcomp}
\usepackage[export]{adjustbox}
\usepackage{float}
\usepackage{booktabs}
\usepackage{multicol}
\usepackage{xparse}
\usepackage{color,soul}
\usepackage[bottom]{footmisc}
\usepackage[binary-units=true]{siunitx}
\DeclareSIUnit{\belmilliwatt}{Bm}
\DeclareSIUnit{\dBm}{\deci\belmilliwatt}
\DeclareSIUnit[per-mode=symbol,per-symbol=p]{\Bps}{\byte\per\second}
\sethlcolor{yellow}
\usepackage{algpseudocode}

\makeatletter
\def\BState{\State\hskip-\ALG@thistlm}
\makeatother

\usepackage{suffix,mathtools}
\DeclarePairedDelimiterX\MeijerM[3]{\lparen}{\rparen}%
{#3\,\delimsize\vert\begin{smallmatrix}#1 \\ #2\end{smallmatrix}}
\newcommand\MeijerG[8][]{%
  G^{\,#2,#3}_{#4,#5}\MeijerM[#1]{#6}{#7}{#8}}

\WithSuffix\newcommand\MeijerG*[7]{%
  G^{\,#1,#2}_{#3,#4}\MeijerM*{#5}{#6}{#7}}

\def\BibTeX{{\rm B\kern-.05em{\sc i\kern-.025em b}\kern-.08em
    T\kern-.1667em\lower.7ex\hbox{E}\kern-.125emX}}

\title{Sensing Assisted Satellite Backhaul with FBL UL and Broadcast DL for Massive IoT
\thanks{This work has received funding from the Horizon 2020
research and innovation staff exchange grant agreement No
101086387,  by Serbian Ministry of Science, Technological Development and Innovation, through the Science and Technological Cooperation program Serbia-China, Research and development project No 00101957 2025 13440 003 000 620 021 and by Danmarks Frie Forskningsfond (DFF) Project
“3D-Twin” under Grant No 4264-00153B.}
}
\author{
\IEEEauthorblockN{Tijana Devaja\IEEEauthorrefmark{1}, Milica Petkovic\IEEEauthorrefmark{2}, Israel Leyva-Mayorga\IEEEauthorrefmark{1}, Dejan Vukobratovi\' c\IEEEauthorrefmark{2}, and \v Cedomir Stefanovi\' c\IEEEauthorrefmark{1}}
\IEEEauthorblockA{
\IEEEauthorrefmark{1}Department of Electronic Systems, Aalborg University, Aalborg, Denmark (\{tde, ilm, cs\}@es.aau.dk)}
\IEEEauthorrefmark{2}Faculty of Technical Sciences, University of Novi Sad, Serbia (\{milica.petkovic, dejanv\}@uns.ac.rs)}% \\
\begin{document}
\maketitle

\begin{abstract}
Massive Internet of Things (IoT) networks operate with short packets whose reliability is limited by finite blocklength (FBL) effects, while remote deployments increasingly rely on Low Earth Orbit (LEO) satellites for backhaul connectivity that is sensitive to atmospheric attenuation. In this paper, we propose a unified end-to-end framework for satellite-assisted massive IoT networks that jointly models uplink FBL random access, sensing assisted satellite backhaul, and worst user broadcast downlink transmission. Uplink reliability is characterized using stochastic geometry, while atmospheric sensing and conservative SNR margins enable FBL-safe backhaul adaptation. Numerical results reveal an optimal uplink access probability due to the tradeoff between spatial reuse and FBL reliability, and show that sensing assisted backhaul margins significantly improve robustness against attenuation uncertainty.
\end{abstract}
\begin{IEEEkeywords}
Finite blocklength, stochastic geometry, LEO satellite backhaul, integrated sensing and communications, massive IoT, broadcast DL.
\end{IEEEkeywords}

%========================================================
\section{Introduction}
%========================================================

Massive IoT networks generate large volumes of sporadic short packets that are decoded at nearby BSs over interference limited random access \cite{lpwan}. Because such packets operate in the finite blocklength (FBL) regime, classical asymptotic Signal-to-Interference-plus-Noise Ratio (SINR) threshold models become inaccurate and must be replaced with coding-rate-reliability analyses that explicitly capture FBL effects \cite{pkd2016,R3,Durisi15}. In parallel, Low Earth Orbit (LEO) satellites increasingly serve as backhaul (BH) for BSs deployed in remote or underserved areas. These Ku/Ka band satellite links are strongly affected by rain attenuation, pointing errors, and other propagation impairments \cite{A1,SatRain}.

Despite significant advances in terrestrial uplink (UL) FBL analysis and satellite resource management, existing works largely consider these components in isolation. UL performance is often evaluated assuming ideal backhaul, while satellite downlink (DL)/BH optimization typically relies on asymptotic-rate models and does not incorporate FBL constraints or the UL induced traffic load.

This motivates the development of a unified end-to-end slotwork that captures
the complete information flow in satellite-assisted massive IoT deployments
with limited or unreliable terrestrial BH (e.g., remote environmental
monitoring, rural smart metering, and disaster-recovery connectivity), namely
\[
\text{IoT device} \;\to\; \text{BS} \;\to\; \text{Satellite} \;\to\; \text{BS} \;\to\; \text{IoT device},
\]
where uplink sensing data are collected at terrestrial base stations and
forwarded through the LEO satellite backhaul, while control information and
feedback (e.g., acknowledgments and configuration updates) are disseminated to
the devices via downlink broadcast transmission.
This model explicitly couples (i) stochastic-geometry UL performance, (ii) sensing assisted satellite BH and (iii) worst user broadcast DL, all under FBL constraints.

To enable such an integrated analysis, we build on and extend several important research directions. FBL channel coding performance has been characterized in seminal works \cite{R3,pkd2016,Durisi15,durisi}, while stochastic-geometry models have been widely used to study interference statistics, random access, and large-scale network behavior \cite{R5,Haenggi12,vaze,R2,Novlan13}. Recent works have examined FBL-oriented UL reliability for large-scale IoT networks \cite{Hesham,Hesham1,tijana}. On the satellite side, extensive studies have addressed rain attenuation modeling and link budgeting \cite{A1,SatRain}. Moreover, the rise of integrated terrestrial–satellite networks and integrated sensing and communication (ISAC) has created opportunities for BH designs that leverage atmospheric sensing \cite{ISAC1,ISAC}. DL reliability constraints, especially the worst user rate limitations in multicast/broadcast satellite transmissions, have also been explored \cite{worstuser,Park20}.

However, prior work generally treats UL reliability, satellite BH adaptation, and DL robustness separately.
Our main contributions are:
\begin{itemize}
  \item We introduce an UL multiple-access sum-rate metric and formulate a sum-rate maximization problem \cite{summax} over the access probability and code rate, constrained by FBL reliability.
  \item We develop a sensing assisted satellite backhaul model that uses rain-attenuation estimates to compute conservative effective SNRs and derive FBL-safe per-hop rates, consistent with ISAC principles \cite{ISAC1,ISAC}.
  \item We model BS broadcast DL using a worst user SNR constraint and incorporate its FBL error probability into an end-to-end reliability expression \cite{worstuser}.
\end{itemize}

% ------------------- ADDED (key insights paragraph) -------------------
Our analysis provides the following key insights. First, due to the tradeoff between spatial reuse and FBL decoding reliability, the uplink admits an optimal access probability that maximizes the successful UL sum rate. Second, sensing uncertainty in the satellite backhaul can induce substantial rate mismatch and outages if rates are selected based only on the raw sensing driven SNR estimates. Third, the proposed conservative BH adaptation based on SNR margins mitigates this uncertainty and yields robust end-to-end performance, while the broadcast DL is fundamentally limited by the worst SNR device and can become the dominant bottleneck if not properly handled.
% ---------------------------------------------------------------------

The rest of the paper is organized as follows. Section~\ref{sec:system_model} introduces the system model, including UL random access, sensing assisted satellite BH, and worst user broadcast DL. Section~\ref{sec:fbl_analysis} derives FBL performance expressions for all hops and defines the UL sum-rate metric. Section~\ref{sec:results} provides numerical results. Section~\ref{sec:conclusion} concludes the paper.

%========================================================
\section{System Model}
\label{sec:system_model}
%========================================================

We consider a hybrid terrestrial-satellite architecture where IoT devices transmit short packets to terrestrial BSs, which forward the packets via LEO satellites to remote BSs as shown in Fig.~1. These BSs then broadcast DL packets to their associated devices. Time is divided into slots. While short packets naturally arise in IoT access, we use a finite-blocklength reliability abstraction across all hops to maintain an end-to-end reliability-based formulation and to capture conservative rate selection under imperfect channel knowledge (in particular, sensing uncertainty in the satellite backhaul). This modeling choice is also aligned with emerging LEO IoT systems that support short-packet technologies (e.g., NB-IoT/LoRa), where both access and satellite segments operate with short frames and stringent reliability requirements.

%----------------------------
\subsection{Network Topology }
%----------------------------

We consider a large-scale massive IoT deployment where
devices and BSs are randomly scattered on the Earth’s
surface. IoT device locations form a homogeneous Poisson
Point Process (PPP) $\Phi_u$ with density $\lambda_u$, and BS
locations form an independent homogeneous PPP $\Phi_b$ with
density $\lambda_b$.

\begin{figure}[!t]
\centerline{\includegraphics[width=3.5in,height=2in]{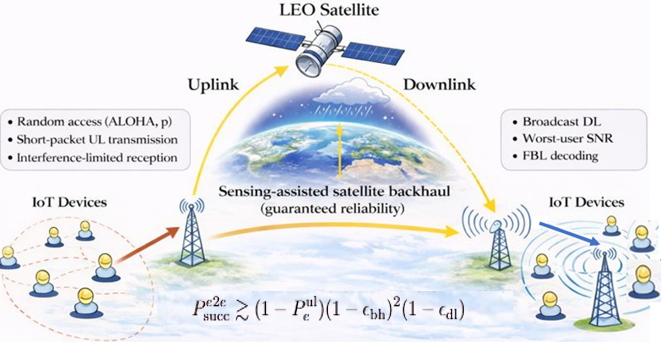}}
\caption{System model.}
\label{Fig_suc1}
\end{figure}

Each device associates to its nearest BS. The distance $R$ from a typical device to its serving BS thus has Probability Density Function (PDF) \cite{R2}
\begin{equation}
  f_R(r)
  = 2\pi \lambda_b\, r\, \exp{\big({-\pi \lambda_b r^2}\big)},  \;\;\;r>0.
\end{equation}

We are assuming normalized unit transmit power and Nakagami
fading environment. The fading power gain $H$ has PDF
\begin{equation}
  f_H(h)
  = \frac{m^m}{\Gamma(m)}\, h^{m-1} \exp {\big({-m h}\big)},\;\; h>0,
\end{equation}
where $m$ is the Nakagami parameter (e.g., $m=1$ reduces to Rayleigh fading).

The path loss for a device at distance $r$ from its BS is modeled as $r^{-\eta}$, with path-loss exponent $\eta>2$. We work in an interference-limited regime where noise is negligible compared to aggregate interference.

Each IoT device generates short information packets of $k$ bits. These packets are encoded into $n^{\rm ul}$ channel uses on the UL, forming a code rate
\begin{equation}
  R_{c}^{\rm {ul}} = \frac{k}{n^{\rm ul}}
  \quad\text{[bits/channel use]}.
  \label{eq:UL_R_def}
\end{equation}
We consider a common information payload of $k$ bits transmitted across all communication phases, while allowing for different blocklengths in the UL, BH, and DL transmissions. Successfully decoded packets at the BS are forwarded over the satellite BH using a blocklength $n^{\rm bh}$, whereas the DL broadcast transmission employs a blocklength $n^{\rm dl}$. Accordingly, the corresponding BH and DL coding rates are given by $R_{c}^{\rm bh} = k / n^{\rm bh}$ and $R_{c}^{\rm dl} = k / n^{\rm dl}$, respectively.

%----------------------------
\subsection{UL IoT-to-BS Random Access}
%----------------------------

In each slot, every IoT device becomes active independently with probability $p$ (slotted ALOHA). Let $a_i\in\{0,1\}$ indicate whether device $i$ is active. The aggregate interference at a reference BS is then
\begin{equation}
  I = \sum_{i\in\Phi_u\setminus\{\text{0}\}} a_i h_i r_i^{-\eta},
  \label{eq:UL_interf}
\end{equation}
where $h_i, r_i$ are the fading power and distance of interfering device $i$ to the reference BS. The index $0$ denotes the typical device and is excluded since its signal is the desired transmission rather than interference.

The instantaneous UL SINR for a typical active device (with distance $R$ and fading $H$ to its nearest BS) is
\begin{equation}
\gamma = \frac{H R^{-\eta}}{I+\sigma_{n}^2} \approx \frac{H R^{-\eta}}{I},
\label{eq2a}
\end{equation}
where $\sigma_{n}^2$ is the background noise power, which is negligible compared to $I$ in the interference limited regime. The random variables $R$, $H$, and $I$ are statistically independent given the PPP model and independent fading across links.

The distribution of $I$ is characterized via its Laplace transform (LT). For PPP devices, ALOHA activity, and Nakagami-$m$ fading, the LT of $I$ has the Kohlrausch-Williams-Watts (KWW) form
\begin{equation}
     \mathcal{L}_I(s)
  = \mathbb{E}[e^{-sI}]
  = \exp\!\Big(-t\, s^{\frac{2}{\eta}}\Big),
\end{equation}
with  $t= p\lambda_u \pi\,
    \frac{\Gamma\!\big(m+\tfrac{2}{\eta}\big)}{\Gamma(m)\,m^{\frac{2}{\eta}}}\,
    \Gamma\!\Big(1-\tfrac{2}{\eta}\Big),
$ shown in \cite{R6}.
The interference PDF $f_I(x)$ is obtained by inverse Laplace transform (ILT). For analytical tractability, and consistently with interference-limited
urban/suburban cellular deployments, we specialize part of the analysis
to the commonly used case $\eta = 4$, where the PDF becomes a Lévy distribution
\begin{equation}
f_{I}(x)=\frac{t \exp\!\big(-\tfrac{t ^2}{4x}\big)}{2\sqrt{\pi}x^{3/2}}.
\label{eq6}
\end{equation}

For later use, we recall the conditional SINR distribution given $h$ and $r$. For $\eta=4$ and the Lévy $f_I$ in \eqref{eq6}, one obtains
\begin{equation}
f_{\Gamma}(\gamma|h,r)
= \frac{t\exp\!\big(-\tfrac{t^2 \gamma}{4h r^{-4}}\big)}{2\sqrt{\pi h r^{-4}\gamma}}.
\label{pdfSINR}
\end{equation}

%----------------------------
\subsection{Satellite Backhaul and Sensing Model}
%----------------------------
We consider a set of LEO satellites indexed by $s\in\mathcal{S}$ and a set of
BSs indexed by $c\in\mathcal{C}$. When satellite $s$ serves BS $c$, the
deterministic free-space path loss is denoted by $L_{s,c}$ and depends on
altitude, elevation angle, and carrier frequency.

In Ka/Ku bands, the dominant random propagation effect is rain attenuation,
modeled by a factor $A_{s,c}\ge 1$ \cite{rain}. The UL BS$\to$satellite segment
experiences both propagation loss and rain attenuation, and its instantaneous
SNR is given by
\begin{equation}
\gamma_{s,c}^{\rm bs}
=\frac{P_c G_c^{\rm tx} G_s^{\rm rx}}
       {L_{s,c}\,A_{s,c}^{\rm ul}\,\ell^{\rm ul}\,\sigma_s^2},
\label{eq:snr_bs_sat}
\end{equation}
where $P_c$ is the BS transmit power, $G_c^{\rm tx}$ and $G_s^{\rm rx}$ denote
the BS transmit and satellite receive gains, respectively, $\ell^{\rm ul}$
captures UL pointing loss, and $\sigma_s^2$ is the satellite noise power.

Similarly, the DL satellite$\to$BS SNR is
\begin{equation}
\gamma_{s,c}^{\rm sb}
=\frac{P_s G_s^{\rm tx} G_c^{\rm rx}}
       {L_{s,c}\,A_{s,c}^{\rm dl}\,\ell^{\rm dl}\,\sigma_c^2},
\label{eq:snr_sat_bs}
\end{equation}
where $P_s$ is the satellite transmit power, $G_s^{\rm tx}$ and $G_c^{\rm rx}$
are the satellite transmit and BS receive gains, $\ell^{\rm dl}$
denotes DL pointing loss, and $\sigma_c^2$ is the BS receiver noise power.
Satellite sensing (e.g., radiometric measurements or pilot-based sounding)
provides an estimate $\widehat{A}_{s,c}$ of the rain attenuation.
We model the sensing process in the logarithmic domain as
\[
\widehat{A}_{s,c}^{(\mathrm{dB})}
= A_{s,c}^{(\mathrm{dB})} + e_{s,c},
\]
where $e_{s,c}\sim\mathcal{N}(0,\sigma^2)$ represents sensing uncertainty and
$\sigma$ characterizes the sensing accuracy \cite{rain}.
Using these attenuation estimates, the corresponding estimated SNRs are
\begin{align}
\widehat{\gamma}_{s,c}^{\rm sb}
&=\frac{P_s G_s^{\rm tx} G_c^{\rm rx}}
        {L_{s,c}\,\widehat{A}_{s,c}^{\rm dl}\,\ell^{\rm dl}\,\sigma_c^2},\\
\widehat{\gamma}_{s,c}^{\rm bs}
&=\frac{P_c G_c^{\rm tx} G_s^{\rm rx}}
        {L_{s,c}\,\widehat{A}_{s,c}^{\rm ul}\,\ell^{\rm ul}\,\sigma_s^2}.
\end{align}

Since attenuation is estimated in the logarithmic domain and mapped to
the linear scale, the resulting SNR estimates are generally biased and
optimistic. While bias correction could compensate for the mean estimation
error, FBL performance is dominated by SNR uncertainty and
lower-tail events rather than average behavior. To ensure reliable
FBL transmission under sensing uncertainty, we introduce
conservative SNR margins $\Delta^{\rm dl}$ and $\Delta^{\rm ul}$ (in dB) and
define the effective SNRs 
\begin{align}
\gamma^{\rm eff,dl}_{s,c}[\mathrm{dB}]
&=\widehat{\gamma}_{s,c}^{\rm sb}[\mathrm{dB}]
  -\Delta^{\rm dl}[\mathrm{dB}],\\[1mm]
\gamma^{\rm eff,ul}_{s,c}[\mathrm{dB}]
&=\widehat{\gamma}_{s,c}^{\rm bs}[\mathrm{dB}]
  -\Delta^{\rm ul}[\mathrm{dB}],
\end{align}
with $\gamma^{\rm eff,dl}_{s,c}$ and $\gamma^{\rm eff,ul}_{s,c}$ denoting the
corresponding linear-scale values. The margins $\Delta^{\mathrm{dl}}$ and $\Delta^{\mathrm{ul}}$ reduce the
probability of rate-selection mismatch events caused by sensing errors,
i.e., cases where the selected coding rate exceeds the instantaneous
channel support. Due to the randomness of the sensing error, a small
residual outage probability may still remain. In this work, the margins
are treated as design parameters that control the tradeoff between
achievable rate and robustness against sensing uncertainty.

%----------------------------
\subsection{Broadcast DL Model}
%----------------------------

% ------------------- ADDED (broadcast motivation) -------------------
We focus on a service model where sensing information collected from remote IoT devices is forwarded to a destination cell and must be disseminated to multiple users that subscribe to the same service (e.g., monitoring/situational awareness). Since the same message is intended for many receivers, broadcast transmission is a natural and resource-efficient choice compared to per-user unicast delivery, and it directly induces a worst-user reliability constraint.
% -------------------------------------------------------------------

Each BS broadcasts a common DL packet per slot (e.g., control message
or aggregated status). The BS does not adapt the rate per user, instead, it
selects a single code rate $R_c^{\rm dl}$ per cell such
that a given target error probability is met at the worst-SNR device. Let $\gamma_{c,u}^{\rm dl}$ denote the DL SNR at device $u$ in cell $c$
under standard power-law path loss and lognormal shadowing
\cite{Haenggi12,Park20}, and $\mathcal{U}_c$ the set of devices
associated with BS $c$. A conservative broadcast SNR is
\begin{equation}
\gamma^{bc}_c = \min_{u \in \mathcal{U}_c} \gamma^{dl}_{c,u},
\label{eq:gamma_bc_def}
\end{equation}
in practice, the SNR at the cell-edge or a low-percentile user.

%========================================================
\section{Finite-Blocklength Performance Analysis}
\label{sec:fbl_analysis}
%========================================================

We now develop FBL reliability expressions for the UL, BH, and broadcast DL, and introduce an UL sum-rate metric. Throughout this paper, reliability refers to the probability of successful packet decoding
under FBL constraints.
In the UL, reliability is determined by the probability that a short packet transmitted by an
active device is decoded correctly at its serving base station.
For the satellite BH and broadcast DL, reliability is imposed through target block error probabilities, which constrain the maximum transmission rates that can be supported under
FBL operation.

%----------------------------
\subsection{UL Block Error Probability}
%----------------------------

For a given SINR value $\gamma$, code rate $R_{\rm ul}$ (bits per channel use), blocklength $n^{\rm ul}$, and target packet error probability $\epsilon$, the normal approximation \cite{R3,pkd2016,Durisi15} gives
\begin{equation}
  \epsilon(\gamma)
  = Q\!\Bigg(
    \sqrt{\frac{n^{\rm ul}}{V(\gamma)}}
    \big(C(\gamma)-R^{\rm ul}_c\big)
  \Bigg),
  \label{eq:UL_fbl_epsilon}
\end{equation}
with $C(\gamma)\log_2(1+\gamma),
  \label{eq:UL_C}$ is the Shannon capacity (bits/channel use), and $V(\gamma)=\gamma\frac{\gamma+2}{(1+\gamma)^2}\log_2^2 e$ is the channel dispersion.

Using the SINR $\gamma$, the unconditional UL block error probability is
\begin{equation}
\begin{split}
P_e^{\rm ul}
&=
\int_{0}^{\infty}\!\epsilon(\gamma)\, f_{\Gamma}(\gamma)\, \rm d\gamma\\
&=
\int_{0}^{\infty}\!\!\int_{0}^{\infty}\!\!\int_{0}^{\infty}
\epsilon(\gamma)\,
f_{\Gamma}(\gamma \mid h,r)\,
f_H(h)\,
f_R(r)\,
\rm d\gamma\, \rm dh\, \rm dr.
\end{split}
\label{eq:UL_Pe_full}
\end{equation}
The conditional SINR distribution $f_{\Gamma}(\gamma|h,r)$ is given by \eqref{pdfSINR}.
Since the integral in \eqref{eq:UL_Pe_full} can not be derived in a closed-form, we
employ the linear approximation of the $Q$-function with the parameters $\mu=\sqrt{\frac{n}{2\pi(2^{2R}-1)\log_2^2e}}$, $\theta=2^R-1$ and $a=\sqrt{\frac{\pi}{2\mu^2}}$. For a short-packet uplink communications, after following derivation in \cite{tijana}, the error
probability is tightly approximated as
\begin{equation}
\begin{split}
P_e^{\rm ul} &\approx \frac{a_1}{\Gamma(m)\pi} \MeijerG*{2}{3}{3}{3}{1, 0, \frac{1}{2}}{\frac{1}{2}, m, 0}{z_1}
+ \frac{a_2}{\Gamma(m)\pi} \MeijerG*{2}{3}{3}{3}{1, 0, \frac{1}{2}}{\frac{1}{2}, m, 0}{z_2} \\
&- b_1 z_1 (c + d z_1^{m-\frac{1}{2}}) + b_2 z_2 (c + d z_2^{m-\frac{1}{2}}),
\end{split}
\label{P_fbl}
\end{equation}
where $a_1 = \frac{1}{2} + \frac{\mu\theta}{\sqrt{2\pi}}, \quad a_2 = \frac{1}{2} - \frac{\mu\theta}{\sqrt{2\pi}}, b_1 = \frac{\mu(\theta + a)}{\sqrt{2\pi}}, \quad b_2 = \frac{\mu(\theta - a)}{\sqrt{2\pi}}, z_1 = \frac{t^2(\theta + a)m}{(\lambda_b \pi)^2}, \quad z_2 = \frac{t^2(\theta - a)m}{(\lambda_b \pi)^2}, c = \frac{\Gamma(m - \frac{1}{2})}{3\sqrt{\pi} \Gamma(m)}, \quad d = \frac{(-1)^m m}{m+1}$,
and $\MeijerG*{\cdot}{\cdot}{\cdot}{\cdot}{\cdot}{\cdot}{\cdot}$ is the Meijer G-function \cite[(9.301)]{grad}. This expression captures the impact of system parameters such as path loss, fading severity, and blocklength, and can be evaluated for performance analysis.

%----------------------------
\subsection{Aggregate UL Throughput and Sum Rate}
%----------------------------

Consider a representative BS cell of area $A_c$. The density of devices inside this cell is $\lambda_u A_c$, and each device transmits in a given slot with probability $p$. For each transmission, the packet is successfully decoded with probability $1 - P_e^{\rm ul}$, where $P_e^{\rm ul}$ is given by  \eqref{P_fbl}.

Hence, the mean number of successfully decoded UL packets per slot at BS $c$ is
\begin{equation}
  \lambda_c^{\rm s}
  = \lambda_u A_c\, p\,(1 - P_e^{\rm ul})
  \quad\text{[packets/slot]}.
  \label{eq:UL_lambda_s}
\end{equation}
Using \eqref{eq:UL_R_def}, the UL sum rate can be defined as the mean number of successfully delivered information bits per slot in cell $c$ defined in \cite{Haenggi12}
\begin{equation}
  R_{\rm sum}^{\rm ul}(c)
  = \lambda_c^{\rm s}\, k
  = \lambda_u A_c\, p\,(1 - P_e^{\rm ul})\, k
  \quad\text{[bits/slot]}.
  \label{eq:UL_sumrate_bits}
\end{equation}
Equivalently, in terms of spectral efficiency,
\begin{equation}
  R_{\rm sum}^{\rm ul}(c)
  = \lambda_u A_c\, p\,(1 - P_e^{\rm ul})\, R^{\rm ul}_c
  \quad\text{[bits/channel use per cell]}.
  \label{eq:UL_sumrate_spectral}
\end{equation}
Equations \eqref{eq:UL_sumrate_bits}–\eqref{eq:UL_sumrate_spectral}
define an mean UL successful throughput (goodput) metric, which
explicitly accounts for random access, interference, and FBL
decoding errors. This metric differs from asymptotic Shannon sum rates and
reflects the effective number of reliably delivered information bits per slot.

%----------------------------
\subsection{Backhaul FBL-Constrained Rates}
%----------------------------

Let $n^{\rm bh}$ denote the BH blocklength and $\epsilon^{\mathrm{bh}}$ the target
block error probability. For any SNR value $\gamma$, the FBL normal
approximation gives the maximum reliable rate \cite{R3}
\begin{equation}
R_{\max}^{\rm FBL}(\gamma,n^{\rm bh},\epsilon^{\mathrm{bh}})
= C(\gamma)
- \sqrt{\frac{V(\gamma)}{n^{\rm bh}}}
  \,Q^{-1}(\epsilon^{\mathrm{bh}}).
\label{eq:RmaxFBL_bh}
\end{equation}
Here, $\epsilon^{\mathrm{bh}}$ denotes the target block error probability for the satellite backhaul link.

Applying \eqref{eq:RmaxFBL_bh} to the effective SNRs of the two hops yields
the FBL-safe UL and DL rates:
\begin{align}
R_{\rm safe}^{\rm ul}(s,c)
&= R_{\max}^{\rm FBL}\!\big(
    \gamma^{\rm eff,ul}_{s,c},\,n^{\rm bh},\,\epsilon^{\mathrm{bh}}
  \big),\\[1mm]
R_{\rm safe}^{\rm dl}(s,c)
&= R_{\max}^{\rm FBL}\!\big(
    \gamma^{\rm eff,dl}_{s,c},\,n^{\rm bh},\,\epsilon^{\mathrm{bh}}
  \big).
  \label{rup}
\end{align}

If nominal code rates $R_{\rm bh}^{\rm ul}$ and $R_{\rm bh}^{\rm dl}$ are
specified for the two hops, the actually usable rates are clipped as
\begin{align}
R^{\rm use,ul}_{s,c}
&=\max\{0,\ \min(R_{\rm bh}^{\rm ul},R_{\rm safe}^{\rm ul}(s,c))\},
\\
R^{\rm use,dl}_{s,c}
&=\max\{0,\ \min(R_{\rm bh}^{\rm dl},R_{\rm safe}^{\rm dl}(s,c))\}.
\end{align}

A packet must traverse both hops successfully. Thus the end-to-end usable
rate of backhaul link $(s,c)$ is the bottleneck value
\begin{equation}
R^{\rm use,e2e}_{s,c}
=\min\!\big\{R^{\rm use,ul}_{s,c},\; R^{\rm use,dl}_{s,c}\big\}.
\label{eq:Ruse_e2e}
\end{equation}

Let $B_s$ be the bandwidth of satellite $s$ and $T_F$ the slot duration. If
a fraction $\rho_{s,c}\!\in\![0,1]$ of the time-bandwidth resource of
satellite $s$ is allocated to BS $c$, the number of useful information bits
delivered end-to-end in that slot is
\begin{equation}
R_{s,c}^{\rm bh}
= \rho_{s,c}\, B_s T_F\, R^{\rm use,e2e}_{s,c}
\quad [\mathrm{bits/slot}].
\label{eq:R_bh_bits_e2e}
\end{equation}
Following standard satellite communication models, the number of information bits delivered in one slot equals the allocated time–bandwidth resource multiplied by the achievable coding rate \cite{Roddy2006}.

%----------------------------
\subsection{Broadcast FBL Reliability and End-to-End Success}
%----------------------------
\begin{figure}[!t]
\centerline{\includegraphics[width=3.5in,height=2.7in]{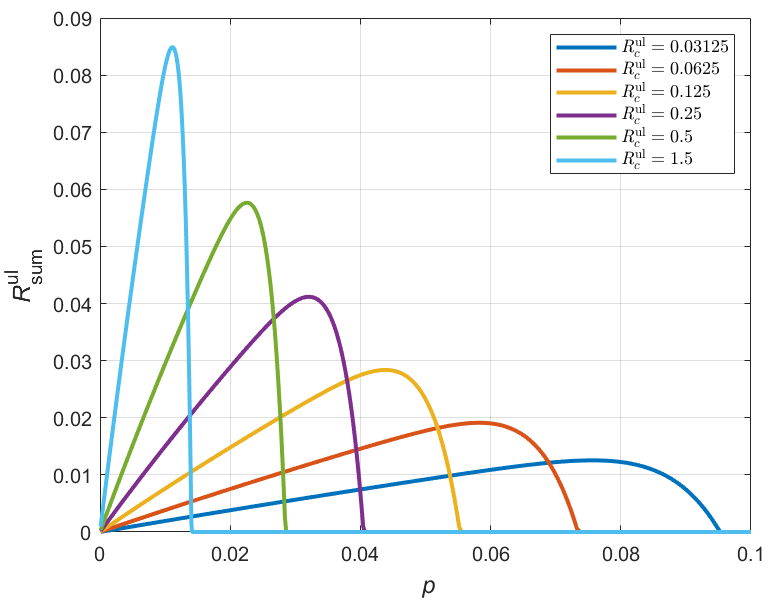}}
\caption{UL sum rate $R_{\rm sum}^{\rm ul}$ versus the access probability $p$ for several code rates $R^{\rm ul}_c$. Densities are $\lambda_b={0.08}$, $\lambda_u={0.5}$, Nakagami-$m$ fading parameters $m =5$ and $n^{\rm{ul}}=128$.}
\label{Fig_suc1}
\end{figure}
For blocklength $n^{\rm dl}$ and rate $R_c^{\rm dl}$, the FBL block error
probability at the edge user with SNR $\gamma_c^{\rm bc}$ in \eqref{eq:gamma_bc_def} is
\begin{equation}
\epsilon_{\rm dl}
= Q\!\Bigg(
    \sqrt{\frac{n^{\rm dl}}{V(\gamma_c^{\rm bc})}}
    \big(C(\gamma_c^{\rm bc})-R_c^{\rm dl}\big)
  \Bigg),
\label{eq:eps_dl}
\end{equation}
with $C(\cdot)$ and $V(\cdot)$ defined earlier.
We require $\epsilon^{\rm dl}_c \le \bar\epsilon^{\rm dl}$ for
all cells $c$. Here, $\bar{\epsilon}^{\rm dl}$ denotes the target block error probability for the broadcast DL.
This yields the maximum admissible broadcast rate
\begin{equation}
R_c^{\rm dl}
\le
R_{\max}^{\rm FBL}\big(\gamma_c^{\rm bc},n^{\rm dl},\epsilon^{\rm dl}\big).
\label{eq:Rdl_max}
\end{equation}

For a given device packet, three reliability events are critical
(i) UL decoding success at the serving BS: probability $1-P_e^{\rm ul}$,
(ii) BH success (BS$\to$satellite$\to$BS) on the selected
route: since each BH hop is designed to satisfy target block error
probability $\epsilon_{bh}$, the two-hop BH success probability is
$(1-\epsilon_{bh})^2$, which is approximately $1-2\epsilon_{bh}$ for
small $\epsilon_{bh}$
(iii) broadcast DL success at the device: probability at least $1-\epsilon^{\rm dl}$.

The expression in \eqref{eq:e2e_succ} serves as a tractable approximation of the
end-to-end packet success probability. It is obtained by assuming statistical
independence among the UL decoding outcome, the satellite BH
transmission, and the broadcast DL reception, and by neglecting
queueing-induced packet drops and excessive delays. Under these assumptions,
the end-to-end success probability can be lower bounded as
\begin{equation}
P^{e2e}_{\mathrm{succ}} \gtrsim (1 - P_e^{\rm{ul}})(1 - \epsilon_{\rm{bh}})^2(1 - \epsilon_{\rm{dl}}),
\label{eq:e2e_succ}
\end{equation}
which explicitly highlights how improvements in UL FBL
reliability, backhaul effective SNR, or broadcast SNR translate into
end-to-end communication reliability.

%========================================================
\section{Numerical Results}
\label{sec:results}
%========================================================

In this section, we present numerical results illustrating the impact of sensing assisted backhaul and worst user broadcast design on the end-to-end reliability of massive IoT networks. For the purpose of numerical evaluation, we assume that the
rain attenuation factor of the satellite BH is identical in
the UL and DL directions, i.e., $A^{ul}_{s,c}=A^{dl}_{s,c}=A_{s,c}$.
This is reasonable when the forward and return links operate over
the same propagation path and within nearby frequency bands, so
that the dominant large-scale rain event affects both directions
similarly. The assumption is adopted here to isolate the impact
of sensing uncertainty and FBL adaptation without
introducing an additional asymmetry between the two BH hops.

Fig.~2 shows the uplink sum rate as a function of the access
probability $p$ for different transmission rates $R_c^{\mathrm{ul}}$
under FBL constraints. For small values of $p$,
the uplink sum rate increases almost linearly as more devices
access the channel. As $p$ grows, the resulting increase in
interference leads to a rapid rise in the decoding error
probability, causing a pronounced peak and subsequent collapse
of the sum rate. This behavior reflects the tradeoff between
spatial reuse and FBL reliability. In particular,
the reliability effect is captured implicitly through the factor
$(1-P_e^{\mathrm{ul}})$ in \eqref{eq:UL_sumrate_spectral}, increasing the access probability $p$ improves spatial reuse by
allowing more devices to transmit simultaneously, but it also
increases aggregate interference, which raises the UL 
error probability $P_e^{\mathrm{ul}}$ under FBL
decoding. 

\begin{figure}[!t]
\centerline{\includegraphics[width=3.5in,height=2.7in]{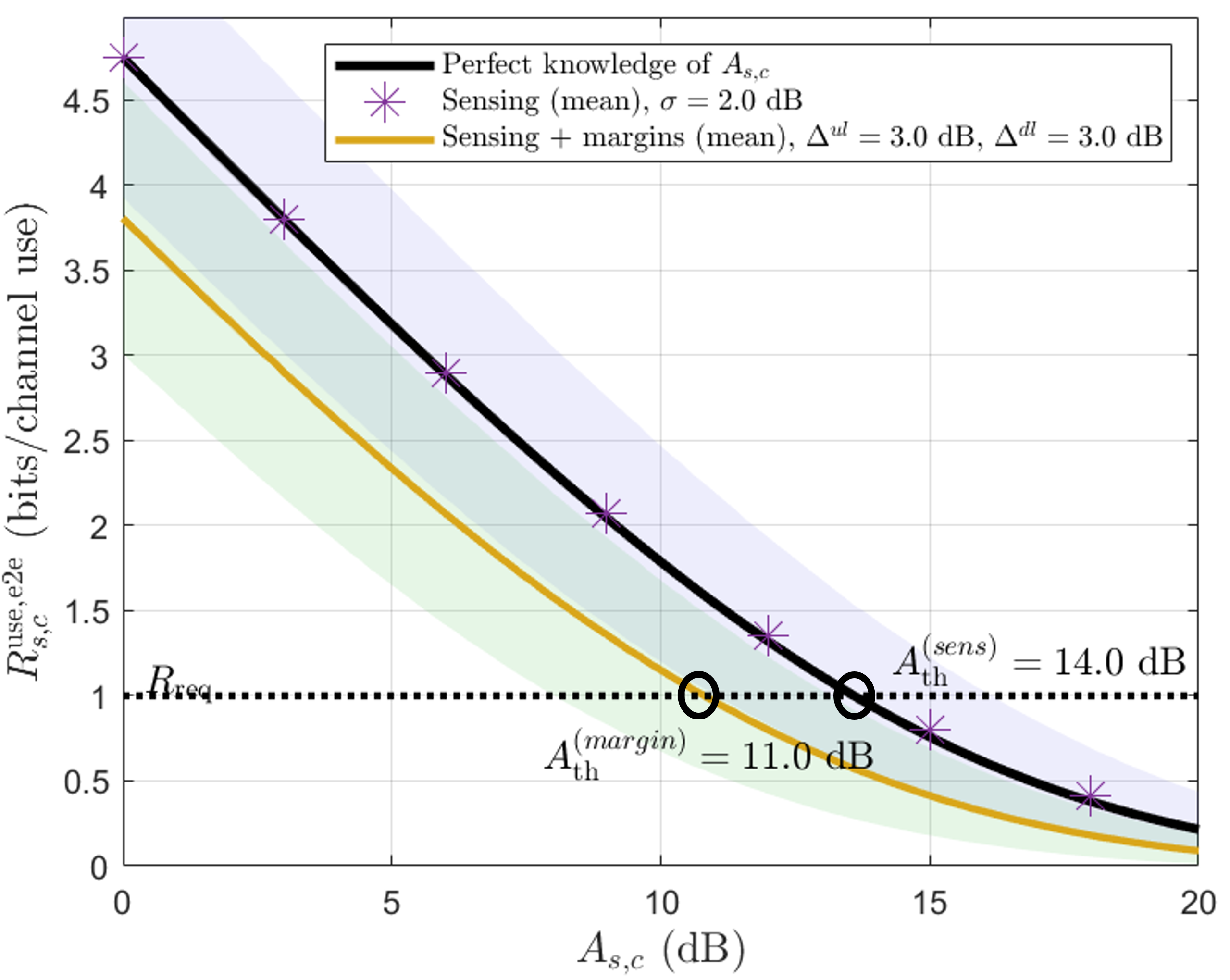}}
\caption{Usable end-to-end satellite BH rate $R^{\mathrm{use,e2e}}_{s,c}$ versus rain attenuation $A_{s,c}$ under finite blocklength transmission with $n^{\mathrm{bh}}=256$ and $\epsilon^{\mathrm{bh}}=10^{-3}$.
Sensing errors have standard deviation $\sigma=2$~dB, and sensing margins $\Delta^{\mathrm{ul}}=\Delta^{\mathrm{dl}}=3$~dB are applied.}
\label{Fig_suc1}
\end{figure}

Fig.~4 illustrates the impact of rain attenuation and sensing uncertainty on the end-to-end satellite backhaul rate $R^{\mathrm{use,e2e}}_{s,c}$ under finite blocklength transmission. As the attenuation $A_{s,c}$ increases, the achievable BH rate decreases monotonically due to the simultaneous degradation of the uplink and downlink SNRs, with the weaker hop determining the end-to-end performance. The case with perfect knowledge of $A_{s,c}$ serves as an upper performance bound.
When sensing-based estimates are used without margins, the mean BH rate closely follows
this upper bound but exhibits non-negligible variability illustrated by the shaded regions caused by
attenuation estimation errors. Importantly, agreement in the mean does not guarantee reliable operation, sensing errors induce rate-selection mismatch events in which the selected coding rate exceeds the instantaneous channel support, leading to non-negligible outage (decoding failure) probability. The conservative margins are introduced precisely to reduce such events.
Introducing conservative sensing margins reduces the achievable rate while significantly improving robustness, leading to a narrower performance spread and a more stable operating region. The attenuation thresholds $A_{\mathrm{th}}^{(\mathrm{sens})}= 14 \rm{dB}$ and $A_{\mathrm{th}}^{(\mathrm{margin})}= 11 \rm{dB}$ quantify the maximum rain attenuation for which the mean backhaul rate remains above the required rate $R_{\mathrm{req}}$, highlighting the fundamental tradeoff between throughput efficiency and reliability in sensing assisted satellite backhaul design. An example of $A_{\mathrm{th}}^{(\mathrm{sens})}$ and $A_{\mathrm{th}}^{(\mathrm{margin})}$ is explicitly indicated in Fig.~3
as the intersection points between the corresponding mean BH rate curves and the required rate $R_{\mathrm{req}}$. The shaded regions represent the variability due to sensing errors, with the wider purple region corresponding to sensing-only operation and the narrower green region corresponding to sensing with conservative margins, highlighting the tradeoff between mean performance and robustness.

Fig.~4 depicts the sensing assisted end-to-end success probability for different uplink blocklengths. In the simulation, the sensing assisted backhaul and broadcast DL error probabilities are fixed to $\epsilon^{\mathrm{bh}}=10^{-3}$ and $\epsilon^{\mathrm{dl}}=10^{-4}$, respectively. As the device density increases, the end-to-end success probability decreases due to increased interference and contention in massive random access. Increasing the uplink blocklength improves reliability, particularly at moderate and high device densities, by alleviating the finite-blocklength penalty on the UL transmission.

\begin{figure}[!t]
\centerline{\includegraphics[width=3.5in,height=2.75in]{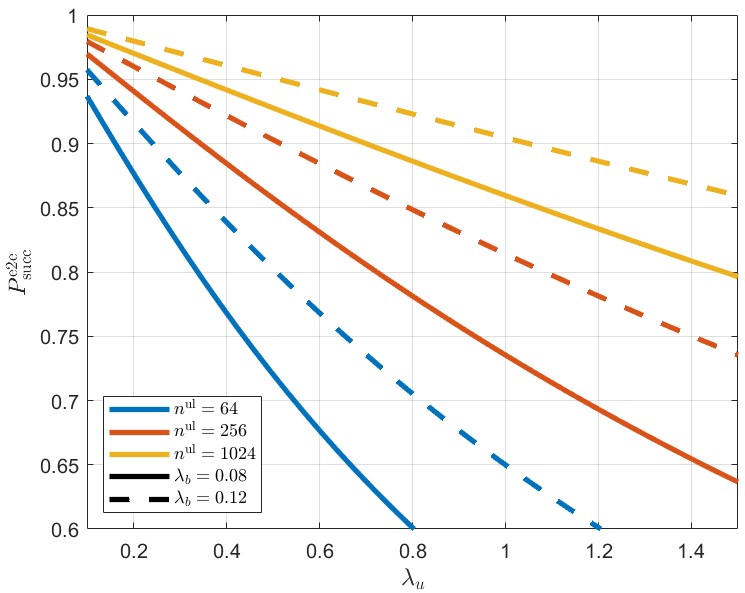}}
\caption{$P_{\mathrm{succ}}^{\mathrm{e2e}}$ as a function of the IoT device density $\lambda_u$ for different uplink blocklengths $n^{\rm {ul}}$. The access probability is fixed to $p=0.05$ and payload is fixed to $k=32$ bits.}
\label{Fig_suc1}
\end{figure}

%========================================================
\section{Conclusion}
\label{sec:conclusion}
%========================================================

We have developed a unified end-to-end model for massive IoT networks that jointly captures UL FBL random access, sensing assisted satellite BH, and worst user broadcast DL operation. Our analysis reveals a fundamental tradeoff between UL spatial reuse and FBL reliability, leading to an optimal access probability that maximizes the UL sum rate. For the satellite BH, we show that sensing margins significantly enhance robustness against attenuation estimation errors at the cost of reduced throughput. Furthermore, the broadcast DL performance is shown to be governed by the worst user SNR, requiring appropriate selection of the DL blocklength and rate to prevent it from becoming the end-to-end bottleneck. Overall, the numerical results demonstrate that coordinated design of UL access, sensing assisted BH, and broadcast DL under FBL constraints is essential to achieving reliable end-to-end communication in massive IoT networks.

%========================================================

\end{document}